\documentstyle[12pt,aaspp4]{article}
\begin{document}
\newcommand{\etal}{{\frenchspacing et al.}}
\newcommand{\ie}{{\em \frenchspacing i.e.}}
\newcommand{\Sub}[2]{\mbox{$#1_{\mbox{\scriptsize #2}}$}}
\newcommand{\Sup}[2]{\mbox{$#1^{\mbox{\scriptsize #2}}$}}
\newcommand{\beq}{\begin{equation}}
\newcommand{\eeq}{\end{equation}}
\newcommand{\lp}{\left(}
\newcommand{\rp}{\right)}

\title{Baryons, Dark Matter, and the Jeans Mass in Simulations of
Cosmological Structure Formation}
\author{J. Michael Owen}
\affil{Department of Astronomy, Ohio State University, Columbus, OH
43210 USA \\
Email:  owen@payne.mps.ohio-state.edu}
\author{Jens V. Villumsen}
\affil{Max Planck Institut f\"{u}r Astrophysik, Karl Schwarzschild Strasse 1,
85740 Garching bei Munchen, Germany \\
Email:  jens@mpa-garching.mpg.de}
\begin{abstract}
We investigate the properties of hybrid gravitational/hydrodynamical
simulations, examining both the numerics and the general physical
properties of gravitationally driven, hierarchical collapse in a mixed
baryonic/dark matter fluid.  We demonstrate that, under certain
restrictions, such simulations converge with increasing resolution to a
consistent solution.  The dark matter achieves convergence provided that
the relevant scales dominating nonlinear collapse are resolved.  If the gas
has a minimum temperature (as expected, for example, when intergalactic gas
is heated by photoionization due to the ultraviolet background) {\em and}
the corresponding Jeans mass is resolved, then the baryons also converge.
However, if there is no minimum baryonic collapse mass or if this scale is
not resolved, then the baryon results err in a systematic fashion.  In such
a case, as resolution is increased the baryon distribution tends toward a
higher density, more tightly bound state.  We attribute this to the fact
that under hierarchical structure formation on all scales there is always
an earlier generation of smaller scale collapses, causing shocks which
irreversibly alter the state of the baryon gas.  In a simulation with
finite resolution we therefore always miss such earlier generation
collapses, unless a physical scale is introduced below which such structure
formation is suppressed in the baryons.  We also find that the baryon/dark
matter ratio follows a characteristic pattern, such that collapsed
structures possess a baryon enriched core (enriched by factors $\sim 2$ or
more over the universal average) which is embedded within a dark matter
halo, even without accounting for radiative cooling of the gas.  The dark
matter is unaffected by changing the baryon distribution (at least in the
dark matter dominated case investigated here), allowing hydrodynamics to
alter the distribution of visible material in the universe from that of the
underlying mass.
\end{abstract}

\section{Introduction}
Hydrodynamics is thought to play a key role in the formation of the visible
structures in the universe, such as bright galaxies and hot intracluster
gas.  For this reason there is a great deal of interest in incorporating
hydrodynamical effects into cosmological structure formation simulations in
order to make direct, quantitative comparisons of such simulations to
observed data.  In addition to gravitation, a cosmological hydrodynamical
simulation must minimally account for pressure support, shock physics, and
radiative cooling, as these are the fundamental physical processes thought
to play a dominant role in the formation of large, bright galaxies (White
\& Rees 1978).  There is already a bewildering array of such studies
published, including Cen \& Ostriker (1992a,b), Katz, Hernquist, \&
Weinberg (1992), Evrard, Summers, \& Davis (1994), Navarro \& White (1994),
and Steinmetz \& M\"{u}ller (1994), to name merely a few.  In order to
appreciate the implications of such ambitious studies, it is important that
we fully understand both the physical effects of hydrodynamics under a
cosmological framework and the numerical aspects of the tools used
for such investigations.  Basic questions such as how the baryon to dark
matter ratio varies in differing structures (galaxies, clusters, and
filaments) and exactly how this is affected by physical processes such as
shock heating, pressure support, or radiative cooling remain unclear.  It
is also difficult to separate real physical effects from numerical
artifacts, particularly given the current limitations on the resolution
which can be achieved.  For example, in a recent study of X-ray clusters
Anninos \& Norman (1996) find the observable characteristics of a simulated
cluster to be quite resolution dependent, with the integrated X-ray
luminosity varying as $L_x \propto (\Delta x)^{-1.17}$, core radius $r_c
\propto (\Delta x)^{0.6}$, and emission weighted temperature $T_X \propto
(\Delta x)^{0.35}$ (where $\Delta x$ is the gridcell size of the
simulation).  In a study of the effects of photoionization on galaxy
formation, Weinberg, Hernquist, \& Katz (1996) find that the complex
interaction of numerical effects (such as resolution) with microphysical
effects (such as radiative cooling and photoionization heating) strongly
influences their resulting model galaxy population.  In this paper we focus
on separating physical from numerical effects in a series of idealized
cosmological hydrodynamical simulations.  This study is intended to be an
exploratory survey of hydrodynamical cosmology, similar in spirit to the
purely gravitational studies of Melott \& Shandarin (1990), Beacom \etal\
(1991), and Little, Weinberg, \& Park (1991).

We will examine the effects of pressure support and shock heating in a
mixed baryonic/dark matter fluid undergoing gravitationally driven
hierarchical collapse.  This problem is approached with two broad questions
in mind: how stable and reliable is the numerical representation of the
system, and what can we learn about the physics of such collapses?  These
questions have been investigated for purely gravitational systems in
studies such as those mentioned above.  In those studies numerically it is
found that the distribution of collisionless matter converges to consistent
states so long as the nonlinear collapse scale is resolved.  Such
convergence has not been demonstrated for collisional systems, however.  It
is not clear that hydrodynamical simulations will demonstrate such
convergence in general, nor if they do that the nonlinear scale is the
crucial scale which must be resolved.  Hydrodynamical processes are
dominated by localized interactions on small scales, allowing the smallest
scales to substantially affect the state of the baryonic gas.  As an
example, consider a collisional fluid undergoing collapse.  Presumably such
a system will undergo shocking near the point of maximal collapse, allowing
a large fraction of the kinetic energy of the gas to be converted to
thermal energy.  In a simple case such as a single plane-wave perturbation
(the Zel'dovich pancake collapse), the obvious scale which must be resolved
is the scale of the shock which forms around the caustic.  However, in a
hierarchical structure formation scenario there is a hierarchy of collapse
scales, and for any given resolution limit there is always a smaller scale
which will undergo nonlinear collapse.  The subsequent evolution of the gas
could well depend upon how well such small scale interactions are resolved,
and changes in the density and temperature of gas on small scales could in
turn influence how it behaves on larger scales (especially if cooling is
important).

In this paper we examine a series of idealized experiments, evolving a
mixed fluid of baryons and collisionless dark matter (dark matter dominated
by mass), coupled gravitationally in a flat, Einstein-de Sitter cosmology.
The mass is seeded with Gaussian distributed initial density perturbations
with a power-law initial power spectrum.  We perform a number of
simulations, varying the resolution, the initial cutoff in the density
perturbation spectrum, and the minimum allowed temperature for the baryons.
Enforcing a minimum temperature for the baryons implies there will be a
minimal level of pressure support, and therefore a minimum collapse scale
(the Jeans mass), below which the baryons are pressure supported against
collapse.  From the numerical point of view, performing a number of
simulations with identical initial physical conditions but varying
resolution allows us to unambiguously identify resolution effects.  By
enforcing a Jeans mass for the baryons we introduce an intrinsic mass scale
to the problem, which may or may not be resolved in any individual
experiment.  The hope is that even if the gas dynamical results do not
converge with increasing resolution in the most general case, the system
will converge if the fundamental Jeans mass is resolved.

The effects of the presence (or absence) of a baryonic Jeans mass also
raises interesting physical questions.  Though we simply impose arbitrary
minima for the baryon temperatures here, processes such as photoionization
enforce minimum temperatures in the real universe by injecting thermal
energy into intergalactic gas.  The Gunn-Peterson test indicates that the
intergalactic medium is highly ionized (and therefore at temperatures $T
\gtrsim 10^4$K) out to at least $z \lesssim 5$.  Shapiro, Giroux, \& Babul
(1994) discuss these issues for the intergalactic medium.  The dark matter,
however, is not directly influenced by this minimal pressure support in the
baryons, and therefore is capable of collapsing on arbitrarily small
scales.  Pressure support provides a mechanism to separate the two species,
and since the dark matter dominates the mass density it can create
substantial gravitational perturbations on scales below the Jeans mass.
While there are many studies of specific cosmological models with detailed
microphysical assumptions, the general problem of the evolution of pressure
supported baryons in the presence of nonlinear dark matter starting from
Gaussian initial conditions has not been investigated in a systematic
fashion.

This paper is organized as follows.  In \S \ref{Sim.sec} we discuss the
particulars of how the simulations are constructed and performed.  In \S
\ref{Numresults.sec} we characterize the numerical effects we find in these
simulations, and in \S \ref{Physresults.sec} we discuss our findings about
the physics of this problem.  Finally, \S \ref{disc.sec} summarizes the
major results of this investigation.

\section{The Simulations}
\label{Sim.sec}
A survey such as this optimally requires a variety of simulations in order
to adequately explore the range of possible resolutions and input physics.
Unfortunately, hydrodynamical cosmological simulations are generally quite
computationally expensive, and therefore in order to run a sufficiently
broad number of experiments we restrict this study to 2-D simulations.
There are two primary advantages to working in 2-D rather than 3-D.  First,
parameter space can be more thoroughly explored, since the computational
cost per simulation is greatly reduced and a larger number of simulations
can be performed.  Second, working in 2-D enables us to perform much higher
resolution simulations than are feasible in 3-D.  While the real universe
is 3-D and we must therefore be cautious about making specific quantitative
predictions based on this work, 2-D experiments can be used to yield
valuable qualitative insights into the behaviour of these systems.  For
similar reasons Melott \& Shandarin (1990) and Beacom \etal\ (1991) also
utilize 2-D simulations in their studies of purely gravitational dynamics.

The 2-D simulations presented here can be interpreted as a slice through an
infinite 3-D simulation (periodic in $(x,y)$ and infinite in $z$).  The
particles interact as parallel rods of infinite length, obeying a
gravitational force law of the form $\Sub{F}{grav} \propto 1/r$.  The
numerical technique used for all simulations is SPH (Smoothed Particle
Hydrodynamics) for the hydrodynamics and PM (Particle-Mesh) for the
gravitation.  The code and technique are described and tested in Owen
\etal\ (1996), so we will not go into much detail here.  We do note,
however, that while our code implements ASPH (Adaptive Smoothed Particle
Hydrodynamics) as described in our initial methods paper, we are not using
the tensor smoothing kernel of ASPH for this investigation, but rather
simple SPH.  The results should be insensitive to such subtle technique
choices since the goal is to compare simulation to simulation, so we employ
simple SPH in order to separate our findings from questions of technique.

All simulations are performed under a flat, Einstein-de Sitter cosmology,
with 10\% baryons by mass ($\Sub{\Omega}{bary}=0.1, \Sub{\Omega}{dm}=0.9,
\Lambda=0$).  Thus the mass density is dominated by the collisionless dark
matter, which is linked gravitationally to the collisional baryons.  The
baryon and dark matter particles are initialized on the same perturbed
grid, with equal numbers of both species.  Therefore, initially all baryons
exactly overlie the dark matter particles, and only hydrodynamical effects
can separate the two species.  The baryon/dark matter mass ratio is set by
varying the particle mass associated with each species.  The initial density
perturbation spectrum is taken to be a power-law $P(k) = \langle |\delta
\rho(k)/ \bar{\rho}|^2 \rangle \propto k^n$ up to a cutoff frequency $k_c$.
Note that since these are 2-D simulations, for integrals over the power
spectrum this is equivalent in the 3-D to a power spectrum of index $n -
1$.  In this paper we adopt a ``flat'' ($n = 0$) 2-D spectrum
\beq
  \Sub{P}{2-D}(k) = \Sub{A}{norm} \left\{ \begin{array}{l@{\quad : \quad}l}
                        k^0 & k \le k_c \quad \Rightarrow \quad 
                                \Sub{P}{3-D}(k) = k^{-1} \\
                        0 & k > k_c, \end{array} \right.
\eeq
where $\Sub{A}{norm}$ normalizes the power-spectrum.  Note that using a
flat cosmology and power-law initial conditions implies these simulations
are scale-free, and should evolve self-similarly in time.  We can choose to
assign specific scales to the simulations in order to convert the
scale-free quantities to physical units.  All simulations are halted after
60 expansion factors, at which point the nonlinear scale (the scale on
which $\delta \rho/\rho \sim 1$) is roughly 1/8 of the box size.

The Jeans length is the scale at which pressure support makes the gas
stable against the growth of linear fluctuations due to self-gravitation --
the Jeans mass is the amount of mass contained within a sphere of diameter
the Jeans length.  The Jeans length $\lambda_J$ and mass $M_J$ are defined
by the well known formula (Binney \& Tremaine 1987)
\beq
  \label{LJ.eq}
  \lambda_J = \lp \frac{\pi c_s^2}{G \rho} \rp^{1/2},
\eeq
\beq
  \label{MJ.eq}
  M_J = \frac{4 \pi}{3} \rho \lp \frac{1}{2} \lambda_J \rp^3
      = \frac{\pi \rho}{6} \lp \frac{\pi c_s^2}{G \rho} \rp^{3/2},
\eeq
where $\rho$ is the mass density and $c_s$ the sound speed.  The baryons
are treated as an ideal gas obeying an equation of state of the form $P =
(\gamma - 1) u \rho$, where $P$ is the pressure and $u$ is the specific
thermal energy.  Enforcing a minimum specific thermal energy (and therefore
temperature) in the gas forces a minimum in the sound speed $c_s^2 = \gamma
P/\rho = \gamma (\gamma - 1) u$, which therefore implies we have a minimum
Jeans mass through equation (\ref{MJ.eq}).  Note that $\rho$ is the total
mass density (baryons and dark matter), since it is the total gravitating
mass which counts, and therefore $M_J$ as expressed in equation
(\ref{MJ.eq}) represents the total mass contained within $r \le
\lambda_J/2$.  If we want the total baryon mass contained within this
radius, we must multiply $M_J$ by $\Sub{\Omega}{bary}/\Omega$.

It is also important to understand how the mass resolution is set for the
baryons by the SPH technique.  This is not simply given by the baryon
particle mass, since SPH interpolation is a smoothing process typically
extending over spatial scales of several interparticle spacings.  In
general the mass resolution for the hydrodynamic calculations can be
estimated as the amount of mass enclosed by a typical SPH interpolation
volume.  If the SPH smoothing scale is given by $h$ and the SPH sampling
extends for $\eta$ smoothing scales, then the mass resolution $M_R$ is
given by
\beq
  \label{MR.eq}
  M_R = \frac{4}{3} \pi (\eta h)^3 \rho.
\eeq
This is probably something of an overestimate, since the weight for each
radial shell in this interpolation volume (given by the SPH sampling kernel
$W$) falls off smoothly towards $r = \eta h$, but given the other
uncertainties in this quantity equation (\ref{MR.eq}) seems a reasonable
estimate.  Note that the resolution limit for the SPH formalism is best
expressed in terms of a mass limit, appropriate for SPH's Lagrangian
nature.  For this reason we choose to express the Jeans limit in terms of
the Jeans mass (eq. [\ref{MJ.eq}]) throughout this work, as the Jeans limit
can be equally expressed in terms of a spatial or a mass scale.  In N-body
work it is common to express the mass resolution of an experiment in units
of numbers of particles.  In our simulations we use a bi-cubic spline
kernel which extends to $\eta = 2$ smoothing lengths, and initialize the
smoothing scales such that the smoothing scale $h$ extends for two particle
spacings.  We therefore have a mass resolution in 2-D of roughly 50
particles, or equivalently in 3-D roughly 260 particles.

We perform simulations both with and without a minimum temperature (giving
Jeans masses $M_J=0$, $M_J>0$), at three different resolutions ($N =
\Sub{N}{bary} = \Sub{N}{dm} = 64^2$, $128^2$, and $256^2$), and for three
different cutoffs in the initial perturbation spectrum ($k_c = 32$, 64, and
128).  The initial density perturbations are initialized as Gaussian
distributed with random phases and amplitudes, but in such a manner that
all simulations have identical phases and amplitudes up to the imposed
cutoff frequency $k_c$.  The cutoff frequencies are the subset of $k_c \in
(32,64,128)$ up to the Nyquist frequency for each resolution $\Sub{k}{Nyq}
= N^{1/2}/2$, so for each resolution we have $k_c(N=64^2) = 32$,
$k_c(N=128^2) \in (32,64)$, and $k_c(N=256^2) \in (32,64,128)$.  For each
value of the minimum temperature we therefore have a grid of simulations
which either have the same input physics at differing resolutions (\ie,
$k_c=32$ for $N \in [64^2, 128^2, 256^2]$), or varying input physics at
fixed resolution (\ie, $N=256^2$ for $k_c \in [32, 64, 128]$).  This allows
us to isolate and study both numerical and physical effects during the
evolution of these simulations.  In total we discuss twelve simulations.

For the simulations with a minimum temperature, there is an ambiguity in
assigning a global Jeans mass with that temperature.  The density in
equation (\ref{MJ.eq}) is formally the {\em local} mass density, and
therefore the Jeans mass is in fact position dependent through
$\rho(\vec{r})$.  Throughout this work we will refer to the Jeans mass at
any given expansion factor as the Jeans mass defined using a fixed minimum
temperature and the average background density, making this mass scale a
function of time only.  This is equivalent to taking the zeroth order
estimate of $M_J$, giving us a well defined characteristic mass scale.  In
terms of this background density, Figure \ref{MJ.fig} shows the baryon
Jeans mass (in units of the resolved mass via equation [\ref{MR.eq}]) as a
function of expansion.  Note that for a given simulation $M_R$ remains
fixed, and it is the Jeans mass which grows as $M_J \propto \rho^{-1/2}
\propto a^{3/2}$.  It is apparent that the $N=256^2$ simulations resolve
the Jeans mass throughout most of the evolution, the $N=128^2$ simulations
resolve $M_J$ by $a/a_i \sim 15$, and the $N=64^2$ simulation does not
approach $M_J/M_R \sim 1$ until the end of our simulations at $a/a_i \sim
60$.  The specific value of $T_{min}$ used in this investigation is chosen
to yield this behaviour.  We discuss physically motivated values for this
minimum temperature in \S \ref{disc.sec}.

\section{Numerical Resolution and the Jeans Mass}
\label{Numresults.sec}
\subsection{Dark Matter}
We will begin by examining the dark matter distribution, as this is a
problem which has been examined previously.  Figures \ref{DMRhoMaps.fig}a
and b show images of the dark matter overdensity
$\Sub{\rho}{dm}/\Sub{\bar{\rho}}{dm}$ for the $M_J=0$ simulations.  In
order to fairly compare with equivalent images of the SPH baryon densities,
the dark matter information is generated by assigning a pseudo-SPH
smoothing scale to each dark matter particle, such that it samples roughly
the same number of neighboring dark matter particles as the SPH smoothing
scale samples in the baryons.  We then use the normal SPH summation method
to assign dark matter densities, which are used to generate these images.
The panels in the figure are arranged with increasing simulation resolution
$N$ along rows, and increasing cutoff frequency $k_c$ down columns.  The
diagonal panels represent each resolution initialized at its Nyquist
frequency for $P(k)$.  Note that the physics of the problem is constant
along rows, and numerics is constant along columns.  If resolution were
unimportant, the results along rows should be identical.  Likewise, since
the numerics is held constant along columns, only physical effects can
alter the results in this direction.

Comparing the dark matter densities along the rows of Figure
\ref{DMRhoMaps.fig}a, it is clear that the structure becomes progressively
more clearly defined as the resolution increases.  This is to be expected,
since the higher resolution simulations can resolve progressively more
collapsed/higher density structures.  The question is whether or not the
underlying particle distribution is systematically changing with
resolution.  In other words, do the simulations converge to the same
particle distribution on the scales which are resolved?  Figure
\ref{DMRhoMaps.fig}b shows this same set of dark matter overdensities for
the $M_J=0$ simulations, only this time each simulation is degraded to an
equivalent $N=64^2$ resolution and resampled.  This is accomplished by
selecting every $n$th node from the higher resolution simulations, throwing
away the rest and suitably modifying the masses and smoothing scales of the
selected particles.  Note that now the dark matter distributions look
indistinguishable for the different resolution experiments, at least
qualitatively.  This similarity implies that the high frequency small scale
structure has minimal effect on the larger scales resolved in this figure.
Looking down the columns of Figure \ref{DMRhoMaps.fig}a it is clear that
increasing $k_c$ does in fact alter the dark matter particle distribution,
such that the large scale, smooth filaments are progressively broken up
into smaller clumps aligned with the overall filamentary structure.  These
differences are lost in the low-res results of Figure \ref{DMRhoMaps.fig}b,
implying that these subtle changes do not significantly affect the large
scale distribution of the dark matter.

In Figures \ref{DMRhoDist.fig}a and b we show the mass distribution
functions for the dark matter overdensity
$f(\Sub{\rho}{dm}/\Sub{\bar{\rho}}{dm})$.  Figure
\ref{DMRhoDist.fig}a includes all particles from each simulation (as in 
Figure \ref{DMRhoMaps.fig}a), while Figure \ref{DMRhoDist.fig}b is
calculated for each simulation degraded to equivalent $N=64^2$ resolutions
(comparable to Figure \ref{DMRhoMaps.fig}b).  The panels are arranged as in
Figure \ref{DMRhoMaps.fig}, with $M_J=0$ and $M_J>0$ overplotted as
different line types.  It is clear that the varying Jeans mass in the
baryons has negligible effect on the dark matter, a point we will return to
in \S \ref{Physresults.sec}.  The full resolution results of Figure
\ref{DMRhoDist.fig}a show a clear trend for a larger fraction of the mass
to lie at higher densities with increasing resolution.  There is also a
similar though weaker trend with increasing $k_c$.  However, examining the
resampled results of Figure \ref{DMRhoDist.fig}b it appears that the
results of all simulations converge, bearing out the visual impressions of
Figures \ref{DMRhoMaps.fig}a and b.  For the dark matter, with increasing
resolution more information is gained about the highest density/most
collapsed fraction of the mass, but so long as the pertinent nonlinear
scales are resolved the results converge.  The underlying particle
distribution does not depend upon the numerical resolution, similarly to
the results discussed in Little \etal\ (1991).

\subsection{Baryons}
We now turn our attention to the baryon distribution.  Figures
\ref{BaryRhoMaps.fig}a, b, c, and d show images of the baryon overdensity for
$M_J=0$ and $M_J>0$ at expansions $a/a_i=30$ and $a/a_i=60$.  There is a
pronounced trend for the collapsed filaments and clumps to become
progressively more strongly defined as the simulation resolution improves
-- even more so than we see in the dark matter.  The tendency to break up
filaments into small scale clumps with increasing $k_c$ is also clearly
evident for the $M_J=0$ case.  Additionally, the presence of a nonzero
Jeans mass visibly influences the baryon density distribution in Figures
\ref{BaryRhoMaps.fig}c and d.  This is particularly evident in the high
resolution $N=256^2$ column, where the increased pressure support creates a
``puffier'' distribution, wiping out the smallest scale structures in the
baryons.  Recall from Figure \ref{MJ.fig} that we naively expect the
presence of the pressure support for $M_J>0$ to affect both $N=128^2$ and
$N=256^2$ at $a/a_i=30$, but not $N=64^2$.  Comparing the results of Figures
\ref{BaryRhoMaps.fig}a and c, we indeed see this trend.  By $a/a_i=60$, the
effects of the Jeans mass are clearly evident (comparing Figures
\ref{BaryRhoMaps.fig}b and d) for $N=128^2$ and $N=256^2$, though $N=64^2$
still appears relatively unaffected.

Figures \ref{BaryRhoMaps_R64.fig}a and b show images of the baryon densities
for the $k_c=32$ simulations, but in this case resampled to $N=64^2$
resolutions analogous to Figure \ref{DMRhoMaps.fig}b.  At expansion $a/a_i
= 30$ (Figure \ref{BaryRhoMaps_R64.fig}a), we see that for $M_J=0$ the
baryons appear to be systematically more tightly collapsed with increasing
simulation resolution, even though they have all been resampled to the same
sampling resolution to produce this image.  This supports the view that the
baryon distribution is fundamentally changing with increasing simulation
resolution, in contrast with the dark matter.  The $N=128^2$ and $N=256^2$
$M_J>0$ simulations, however, demonstrate very similar baryon density
images, though $N=64^2$ still appears different at $a/a_i=30$.  At
$a/a_i=60$ (Figure \ref{BaryRhoMaps_R64.fig}b) we again see for $M_J=0$ a
clear trend with simulation resolution, while the $M_J>0$ runs look
remarkably similar to one another.

Figures \ref{BaryRhoDist.fig}a and b show the full resolution mass
distribution functions of the baryon overdensities
$f(\Sub{\rho}{bary}/\Sub{\bar{\rho}}{bary})$ for all simulations at
$a/a_i=30$ and $a/a_i=60$, respectively.  The $M_J=0$ functions show a
strong trend to transfer mass from low to high densities with increasing
resolution, and a similar though weaker trend with $k_c$.  However, even at
full resolution the $M_J>0$ simulations show very similar density
distributions once $M_J$ is resolved.  The $M_J>0$ simulations also appear
to be relatively insensitive to $k_c$, suggesting that the increased small
scale power is being wiped out by the pressure support.  Figures
\ref{BaryRhoDist.fig}c and d show these same baryon density distribution
functions, only for all simulations degraded to $N=64^2$ resolutions.
These bear out our previous observations.  In the case with no Jeans mass,
there is no sign of convergence in the baryon distribution as the
resolution is increased.  However, when a Jeans mass is present, then the
baryon distributions do converge {\em once the Jeans mass is resolved}.

Figure \ref{KS.fig} presents a more quantitative way to measure this
convergence problem.  In this figure we calculate the Kolmogorov-Smirnov
statistic $D(\Sub{\rho}{bary})$, comparing the baryon density distribution
for each simulation to the others at the same expansion and Jeans mass.  We
do not expect these simulations to exactly reproduce one another, and
therefore there is little point in assigning significance to the exact
quantitative value of $D$.  However, the K-S statistic does provide
objective measures of how similar or dissimilar these distributions are, and
therefore we might expect to learn something by comparing their relative
values.  Comparing the upper panels of Figure \ref{KS.fig} we can see that
at $a/a_i=30$ the $N=128^2$ and $N=256^2$ simulations are more similar for
$M_J>0$ than for $M_J=0$, while the $N=64^2$ simulation remains relatively
distinct in both cases.  At $a/a_i=60$, however, we can see that for
$M_J>0$ all the simulations appear comparable, while for $M_J=0$ they
remain distinct for the different resolutions.

We therefore have a subtly different picture for the numerical behaviour of
the dark matter and baryons.  The critical resolution scale for the dark
matter is the scale of nonlinearity.  So long as this scale is resolved,
the dark matter distribution can be expected to converge to a consistent
state on resolved scales.  Unfortunately, the distribution and state of the
baryonic particles appears in general to be sensitive to the numerical
resolution.  However, it is possible and physically plausible to define a
fundamental collapse scale in the form of the Jeans mass for the baryons,
below which baryonic structure formation is suppressed.  This scale can now
be treated as the critical baryonic resolution scale, and we do find that
once this threshold is reached the baryon distribution will reliably
converge as well.

\section{Hydrodynamics and the Baryon Distribution}
\label{Physresults.sec}
\subsection{Shocks and Temperatures}
The results of the previous section indicate that hydrodynamical
interactions on small scales can significantly alter the the final state of
the baryons in ways which propagate upward and affect larger scales.  The
tendency for a simulation with a given finite resolution is to
underestimate the ``true'' fraction of high density, collapsed baryons.  A
likely cause for this trend is the presence of small scale, unresolved
shocks in the baryon gas.  Because shocks provide a mechanism for
transferring the gas's kinetic energy to thermal energy, it is reasonable
to expect that the fashion and degree to which the baryons collapse will be
dependent upon when and how strongly they undergo shocks.  In this section
we investigate the thermal state of the baryons, with the goal of
understanding the pattern and importance of shocking in the gas.

In the top row of Figure \ref{RhoTDist.fig} we show the 2-D mass
distribution function of the baryons in terms of their overdensity and
temperature $f(\Sub{\rho}{bary}/\Sub{\bar{\rho}}{bary},T)$ for the $M_J=0$
simulations at $a/a_i=60$.  The various resolutions share some gross
properties in the $\rho-T$ plane.  The low density gas tends for the most
part to be cool, though there is a tail of low density material with
temperatures up to $T \lesssim 10^4$K.  The high density gas is at
relatively high temperatures, with most of the material near $T \sim
10^6$K.  However, there is a notable trend for the highest density material
to be somewhat cooler with increasing simulation resolution.  This effect
is similar to the behaviour seen in simple 1-D collapse such as the
Zel'dovich pancake (Shapiro \& Struck-Marcell 1985).  The highest density
gas is the fraction which collapses earliest, when the background density
is highest.  Such gas is placed on a lower adiabat than gas which falls in
at later times, and thus remains cooler.  In our case this means that since
higher resolution simulations can resolve higher density clumps (which
therefore form at earlier times), we should tend to see the temperature of
the highest density material fall with increasing resolution.

The high temperature gas is heated by shocks as it falls into the dark
matter dominated potential wells.  In order to isolate shock heating from
simple adiabatic compression heating, we calculate the distribution of the
temperature in units of the adiabatic temperature $\Sub{T}{ad}$, given by
\beq
  \Sub{T}{ad} = T_0 \lp \frac{\Sub{\rho}{bary}}{\rho_0} \rp^{\gamma - 1}.
\eeq
\Sub{T}{ad} represents the temperature the gas would be at if it were only
heated through simple $P dV$ work.  Since the only non-adiabatic process we
allow is shock heating, only gas which has undergone shocking should be at
$T/\Sub{T}{ad} > 1$.  In the bottom row of Figure \ref{RhoTDist.fig} we
calculate the distribution
$f(\Sub{\rho}{bary}/\Sub{\bar{\rho}}{bary},T/\Sub{T}{ad})$ for the $M_J=0$
simulations at $a/a_i=60$.  The high density fraction of the gas is clearly
strongly shocked in all cases, with $T/\Sub{T}{ad} \sim 10^7-10^9$.  There
is a clear trend for $T/\Sub{T}{ad}$ in the high density gas to fall with
resolution, indicating that the highest density fraction of the gas is less
strongly shocked as the resolution increases.  Though we do not show the
results at fixed resolution and increasing $k_c$ here, there are also
subtle trends evident with $k_c$ in both the $\rho-T$ and
$\rho-T/\Sub{T}{ad}$ planes.  Generally the temperature/shocking
distribution of moderately overdense material grows wider with increasing
$k_c$.

It appears that shocks are indeed the key physical mechanism distinguishing
the different resolution experiments.  We find that in general most of the
baryonic material is processed through shocks at some point.  We note a
general pattern in which the highest density gas in low resolution
experiments is characteristically more strongly shocked than the highest
density gas in higher resolution experiments.  The physical inference of
these trends is that the larger the region which collapses, the stronger
the resulting shock.  The underlying physical mechanism for this property
is easily understood.  Since the highest density material represents the
gas which collapses earliest, this is also the gas which falls into the
shallowest potential wells.  As the structure continues to grow, these
potential wells deepen.  Gas which infalls at later times therefore picks
up more kinetic energy, which in turn leads to stronger shocking and higher
temperatures.  Once shocking occurs, the state of the baryon gas is
discontinuously and irreversibly altered.  In order to properly represent
the physical state of the gas, a simulation must resolve the smallest
scales on which shocks are occuring.  This is why enforcing a Jeans mass
allows convergence, since establishing a minimum Jeans mass implies there
is a minimum scale on which baryonic structures can form, forcing a minimum
scale for shocking.

\subsection{Comparing the Baryon \& Dark Matter Distributions}
One of the most fundamental questions we can address is how the
distributions of dark matter and baryons compare to one another.
Comparing the dark matter and baryon density fields for the $M_J=0$ case in
Figures \ref{DMRhoMaps.fig}a and \ref{BaryRhoMaps.fig}b, there is a
distinct impression that the baryons tend to be more tightly clustered than
the dark matter on all collapsed scales.  The situation is a bit more
complex for the $M_J>0$ case in Figure \ref{BaryRhoMaps.fig}d.  Comparing
the $(N=256^2,k_c=128)$ distributions, it is evident that the $M_J>0$
baryons show a more diffuse structure than that of the $M_J=0$ case, to the
point that some of the smallest scale structures are entirely suppressed.
Bear in mind that the dark matter evolves essentially independently of the
baryons in this dark matter dominated case, so the small scale structures
still form in the overall mass distribution -- the baryons are simply
excluded from them.  The large scale structures such as the filaments and
the largest knots are still quite prominent in the $M_J>0$ baryon
distribution, just as for the $M_J=0$ case.  These patterns suggest that
the baryons are generically more clustered than the dark matter, down to
the scale set by the Jeans mass.  At this scale and lower, the dark matter
continues to form collapsed structures, whereas the baryons are held out of
these structures by the pressure support enforced by the minimum
temperature.

In Figure \ref{B2D.fig} we calculate the baryon to dark matter number
density ratio as a function of baryonic overdensity.  The baryon to dark
matter ratio is defined as $\Sub{n}{bary}/\Sub{n}{dm} =
\Sub{\Omega}{dm}\Sub{\rho}{bary}/\Sub{\Omega}{bary}\Sub{\rho}{dm}$, so that 
$\Sub{n}{bary}/\Sub{n}{dm} > 1$ corresponds to baryon enrichment, while
$\Sub{n}{bary}/\Sub{n}{dm} < 1$ implies baryon depletion.  There is a clear
trend for the highest density material to be baryon enriched, implying that
the cores of the most collapsed structures are relatively enriched in
baryons compared with the universal average.  This trend persists even in
the $M_J>0$ simulations, though it is not as pronounced as in the $M_J=0$
case.  There is no evidence that a significant fraction of the baryons
exist in regions which are dark matter enhanced.  In all simulations
underdense material appears to lie near the universal average mixture
$\Sub{n}{bary}/\Sub{n}{dm} \sim 1$.  We also note a trend with resolution,
such that the higher the resolution of the simulation, the greater the
baryon enrichment found in overdense regions.

A simple physical picture can account for these trends.  So long as the
density evolution is in the linear regime ($\delta \rho/\rho \ll 1$), the
dark matter and baryons evolve together, remaining at the universal mix of
$\Sub{n}{bary}/\Sub{n}{dm} \sim 1$.  During this linear phase the pressure
support (barring any imposed minimum pressure) is orders of magnitude less
important than the gravitational term, so the baryon/dark matter fluid
evolves as a pressureless gas.  Once nonlinear collapse begins ($\delta
\rho/\rho \gtrsim 1$), the baryons rapidly fall inward with the dark matter
until they collide near the potential minimum.  At this point the baryon
gas shocks, converting the majority of its kinetic energy into thermal
energy, and it stops, forming a hot pressure supported gas at the bottom of
the potential well.  In the case with a minimum pressure support, the
collapse proceeds until the pressure term (due to the increase in density)
builds sufficiently to impede the baryons infall, at which point the
baryons slow, separate from the infall, and shock.  In either case the dark
matter forms a more diffuse structure supported by velocity dispersion.
This process leads to the generic patterns noted above: on scales below
which the collapse has become nonlinear, the baryons tend to be
characteristically more clustered than the dark matter, at least down to
the minimal point set by the Jeans scale.  In either case the critical
factor determining exactly when the baryons separate from the general
inflow is the point at which shocking sets in.  We also know from the
numerical observations that this process is resolution dependent, and in
fact the baryon enrichments we see for the $M_J=0$ case in Figure
\ref{B2D.fig} must represent lower limits to the ``true'' baryon
enrichment.  The enrichments noted for the $M_J>0$ simulations should be
reliable, to the extent that the specific minimum temperature chosen is
reasonable.

It is somewhat puzzling to note that our measured positive biasing of the
baryons in collapsed structures is at odds with previously published
results.  In a study of the cluster formation under the standard $\Omega=1$
Cold Dark Matter (CDM) model, Evrard (1990) finds that while outside of the
cluster environment the baryons and dark matter simply track the universal
average mix, the baryon fraction within the cluster is in fact somewhat
lowered.  Kang \etal\ 1994 examine a larger volume of an $\Omega=1$ CDM
cosmology, and find that not only are the overdense regions baryon
depleted, but that their underdense, void structures are baryon enriched.
There are several possible explanations for this disagreement.  One
possibility is that this represents a geometric effect, in that our
experiments are 2-D, while these other studies employ fully 3-D
simulations.  In our simulations, the ``filaments'' actually represent
walls, and the most collapsed knots are best interpreted as cross-sections
through tubular filaments.  The processes of collapsing to a plane, a line,
or a point are certainly different processes, and the isotropy of pressure
support makes these structures progressively more difficult to form.  In a
1-D planar collapse, for instance, it is well known that the central
collapse plane will be baryon enriched, while the question of whether or
not a cluster is baryon enriched or depleted is still hotly debated.  We
see some evidence for this effect in Figure \ref{B2D.fig}.  Looking
particularly at the upper dashed lines in this figure (representing the
baryon enrichment at which 90\% of the mass at that overdensity lies below)
we note our most extreme enrichments occur at moderate overdensities,
roughly in the range $\Sub{\rho}{bary}/\Sub{\bar{\rho}}{bary} \sim
10^1-10^2$.  This extremely baryon enriched material represents the
``filaments'' in our simulations (walls in 3-D).  It is also possible that
resolution effects play a role here.  As pointed out previously, we find a
strong resolution dependence, such that finite resolution tends to
underestimate the fraction of high density, collapsed baryonic material.
Evrard (1990) uses $16^3$ SPH nodes to represent his baryon component,
which for the scale of his box is equivalent to our lowest resolution
simulations.  Kang \etal\ (1994) use an entirely different technique to
simulate the hydrodynamics, which relies upon a fixed grid to represent the
baryons.  This limits their spatial resolution so that typical clusters are
only a handful of cells across.  It is also important to compare these
quantities in the same manner.  In Figure \ref{B2D.fig} we calculate the
baryon to dark matter mixture in a manner which follows the baryon mass,
since we sample at the positions of the baryon particles.  This naturally
gives the greatest weight to the most prominent baryonic structures.  Kang
\etal\ (1994) calculate this distribution in a manner which is volume
weighted, which will tend to give the greatest weight to underdense, void
like regions.  Since the baryon fraction appears to be a function of
environment, these differences can be significant.  Without further study,
it is difficult to know the true reason for this discrepancy, or how the
actual baryon/dark matter ratio should evolve.

\section{Discussion}
\label{disc.sec}
The results of this investigation can be broken into two broad categories:
what is revealed about the physics of hierarchical collapse in a mixed
baryonic/dark matter fluid, and what is learned about the numerics of
simulations of this process.  We find that the dark matter converges to a
consistent state on resolved scales, so long as the nonlinear collapse
scales are well resolved.  Increasing the resolution of the experiment does
not fundamentally alter the dark matter distribution, but simply yields
more detailed information about the small scale collapsed structures.  This
is in agreement with previous, purely collisionless studies, though we
demonstrate this here including a collisional component.

The numerical story is quite different for the collisional baryonic gas.
We find that in the case where we do not impose a fundamental physical
resolution scale in the baryons, the simulation results do not converge
with increasing resolution.  Rather, as the numerical resolution of the
experiment is increased, the collapsed fraction of the baryons is
systematically altered toward a higher density, more tightly bound, and
less strongly shocked state.  The physical reason for this behaviour is the
presence of shocks, which allows the evolution on small scales to affect
the overall state of the baryonic mass.  With improving resolution the
simulation is able to resolve the collapse of smaller structures at earlier
times.  The smaller scale (and therefore earlier) the resolved collapse,
the weaker the resulting shock is found to be.  This effect is most obvious
in Figure \ref{RhoTDist.fig}, where there is a systematic trend of higher
density/more weakly shocked material with increasing resolution.

The fact that the dark matter converges in general with resolution, while
the baryons do not, highlights a fundamental difference in the physics of
these two species.  While both dark matter and baryon fluids react to the
global and local gravitational potential, the baryons are additionally
subject to purely local hydrodynamical phenomena -- most prominently
shocking in this case.  Once strong shocking sets in these hydrodynamical
effects can rise to rival the gravitational force on the baryonic fluid,
allowing the baryons to be strongly influenced by interactions on small
scales in ways which the dark matter is not.  This implies that such
small scale interactions can be just as important as the large scale forces
in determining the final state of the baryons.  In other words, for the
dark matter there is no back reaction from small to large scales, whereas
the baryons are strongly influenced by interactions on small scales.  In
the coupling of these physical processes, gravitation dominates the large
scale structure, but hydrodynamics affects the local arrangements and
characteristics of the baryonic gas.  If we want the quantitative results
of such studies to be reliable, we must have reason to believe that the
localized hydrodynamical processes are adequately resolved.

This gloomy picture is alleviated by an important physical effect: the
Jeans mass.  Introducing a minimum temperature (and therefore pressure
support) into the baryons creates a fundamental length/mass scale, below
which the baryons are supported by pressure against any further collapse or
structure formation.  We find that once we introduce such a minimal scale
into the baryonic component, the simulation results converge as this scale
is resolved.  This convergence holds even though the dark matter component
continues to form structures below the baryon Jeans scale.  Although the
Jeans scale is dependent upon the local density, we find that the global
Jeans scale defined using the background density is adequate to define the
critical resolution necessary for the hydrodynamics to converge.  This
therefore describes an additional resolution scale necessary for
hydrodynamical simulations to meet, much as the nonlinear mass scale
represents the crucial resolution necessary for purely gravitational
systems.  Furthermore, our experiments indicate that equation (\ref{MR.eq})
is a reasonable estimate of an SPH simulation's true mass resolution, since
we find that the threshold $M_R \lesssim M_J$ marks the point at which
convergence is achieved.

In these experiments we have tested the effects of the Jeans mass in an
idealized framework by simply imposing an arbitrary minimum temperature
into our system, but there is reason to believe that such minimum
temperatures should exist in the real universe.  Based upon observations
such as the Gunn-Peterson test (Gunn \& Peterson 1965), it is known that
the IGM is highly ionized out to redshifts $z \lesssim 5$, which implies a
minimum temperature for the IGM of at least $T \gtrsim 10^4$.  Assuming an
Einstein-de Sitter cosmology, a minimum temperature of $T \sim 10^4$
requires a minimum spatial resolution (via eq. [\ref{LJ.eq}])
\beq
  \lambda_J \sim 0.777 \; (1 + z)^{-3/2} \;
  \lp \frac{\mu}{0.6} \rp^{-1/2} \; \lp \frac{T}{10^4 \mbox{K}} \rp^{1/2}
  \; h^{-1} \mbox{\ Mpc},
\eeq
which equates to a baryon mass resolution of (eq. [\ref{MJ.eq}])
\beq
  M_R \lesssim \Sub{\Omega}{bary} \; M_J \sim 
    6.82 \times 10^{10} \; \Sub{\Omega}{bary} \;
    (1 + z)^{-3/2} \; \lp \frac{\mu}{0.6} \rp^{-3/2} \;
    \lp \frac{T}{10^4 \mbox{K}} \rp^{3/2} \; h^{-1} \ M_{\sun}.
\eeq
This limit can also be expressed in terms of a minimum circular velocity,
which has the advantage of being independent of redshift.  The minimum
circular velocity can found as a function of the minimum temperature by
relating the kinetic energy necessary for dynamical support to the internal
energy for equivalent pressure support (Thoul \& Weinberg 1996), yielding
\beq
  \Sub{v}{circ} = \lp \frac{2 k T}{\mu m_p} \rp^{1/2}
    \sim 16.6 \lp \frac{\mu}{0.6} \rp^{-1/2} \lp \frac{T}{10^4 \mbox{K}}
       \rp^{1/2} \mbox{km/sec}.
\eeq
In our $M_J>0$ simulations if we choose to call the scale at which RMS mass
fluctuation is $\Delta M/M \sim 0.5$ to be 8 $h^{-1}$ Mpc at the final
expansion, then our box scale is $L=64 h^{-1}$Mpc and the minimum
temperature corresponds to $\Sub{T}{min} \sim 10^6$K.  While there are some
suggestions that the intergalactic medium could be heated to temperatures
as hot as $10^6$K (through mechanisms such as large scale shocks of the
IGM), clearly these simulations do not meet our criteria if we wish to
consider photoionization as setting the minimum temperature.  It is also
not clear that the current generation of large-scale hydrodynamical
cosmological simulations meet this criterion, but it should be achievable.

It is still unclear whether or not in the case with no minimum temperature
imposed the baryon distribution will eventually converge.  It is well known
that in a purely gravitational system, as structure builds and smaller dark
matter groups merge into larger structures, the dark matter ``forgets''
about the earlier small scale collapses as such small structures are
incorporated into larger halos and disrupted.  This is why the dark matter
results converge once the nonlinear mass scale is resolved.  While it is
evident from studies such as this that the baryons maintain a longer memory
of their previous encounters, it seems likely that as the baryon gas is
progressively processed through larger scale and stronger shocks, at some
point the previous evolution should become unimportant.  At exactly what
level this transition is reached remains uncertain, however, as we see no
evidence for such convergence here.

Radiative cooling must be accounted for in order to model processes such as
galaxy formation, and the inclusion of radiative cooling can only
exacerbate the non-convergence problems we find here.  The amount of energy
per unit mass dissipated by radiative cooling is proportional to the
density, and we have already noted that the trend with finite resolution is to
underestimate the local gas density and overestimate the temperature.
Given these tendencies, it is not difficult to envision problems for finite
resolution simulations which will tend to underestimate the effectiveness
of radiative cooling in lowering the temperature (and therefore pressure
support) of the shocked gas.  This could lead to perhaps drastic
underestimates of the fraction of cold, collapsed baryons for a given
system, and therefore strongly influence the inferred galaxy formation.
Evrard \etal\ (1994) note this effect when comparing their high and low
resolution 3-D simulations.  They find that altering their linear
resolution by a factor of two (and therefore the mass resolution by a
factor of eight) changes the measured total amount of cold collapsed
baryons by a factor of $\sim 3$.  They attribute this change to just the
sort of problems we discuss here.  Weinberg, Hernquist, \& Katz (1996)
report similar findings and interpretation for simulations with a
photoionizing background.  It therefore seems likely it is all the more
important to resolve the minimum mass scale set by the minimum temperature
in systems with radiative cooling.

We find that the majority of the baryonic mass undergoes strong shocking so
long as the nonlinear mass scale exceeds the Jeans mass.  At infinite
resolution in the $M_J=0$ case, it is possible that all of the baryonic
material undergoes shocking.  As anticipated from previous investigations,
the highest density collapsed fraction is characteristically less shocked
as compared with later infalling material from larger regions.  The
underlying cause for this behavior is the fact that potential wells deepen
as structure grows.  The highest density material is that which collapses
earliest due to the smallest scale perturbations.  This material falls into
relatively shallow potential wells, and is only weakly shocked.  As the
structures continue to grow, progressively larger scales go nonlinear and
collapse.  The potential wells deepen and infalling material gains more
energy, resulting in stronger shocking and higher temperatures.

Hydrodynamics can also play an important role in determining the
distribution of the baryon mass, particularly in collapsed structures.  In
the absence of external mechanisms to heat the baryons (such as energy
input from photoionization), during the linear phase of structure growth
the baryons evolve as a pressureless fluid and simply follow the dominant
dark matter.  Once nonlinear collapse sets in, the baryons fall to the
potential minimum, shock, convert their kinetic energy to thermal energy,
and settle.  In contrast, the dark matter simply passes though the
potential minimum and creates a more diffuse structure supported by the
anisotropic pressure of random velocities.  This difference gives rise to a
characteristic pattern in the baryon/dark matter ratio.  Wherever the
evolution is still linear, the baryons and dark matter simply remain at the
universal mix.  With the onset of nonlinear collapse, the baryons fall to
the minimum of the potential well where they form a baryon enriched core,
surrounded by a dark matter rich halo.  We find that even in the absence of
radiative cooling the cores of collapsed structures can become baryon
enriched by factors of $\Sub{n}{bary}/\Sub{n}{dm} \sim 2$ or more, though
this value is likely resolution and dimension dependent.  If the thermal
energy of the baryons is raised to the point that it rivals the potential
energy during the collapse, the baryons will become pressure supported and
stop collapsing at that point.  In all cases we find that the dark matter
is relatively unaffected by the baryon distribution.  This is due to the
fact that the dark matter dominates the mass density, and therefore the
gravitational potential.  In general it appears that under a dark matter
dominated scenario hydrodynamics can substantially alter the
characteristics of the baryonic material (and therefore the visible
universe), such that it does not directly follow the true mass distribution
which is dominated by the the dark matter.

\acknowledgements
We would like to thank David Weinberg for inspiring this project, and for
many useful discussions during its course as well.  We would also like to
thank the members of Ohio State's Astronomy Department for the use of their
workstations both for performance and analysis of many of the simulations.
JMO acknowledges support from NASA grant NAG5-2882 during this project.
Some of these simulations were performed on the Cray Y/MP at the Ohio
Supercomputer Center.

\clearpage
\section{References}
\begin{description}
\item Anninos, P., \& Norman, M. L. 1996, \apj, 459, 12

\item Binney, J. \& Tremaine, S. 1987, {\em Galactic Dynamics},
(Princeton: Princeton University Press)

\item Beacom, J. F., Dominik, K. G., Melott, A. L., Perkins, S. P., \&
Shandarin, S. F. 1991, \apj, 372, 351

\item Cen, R., \& Ostriker, J. 1992a, \apj, 393, 22

\item Cen, R., \& Ostriker, J. 1992b, \apj, 399, L113

\item Evrard, A. E.\ 1990, \apj, 363, 349

\item Evrard, A. E., Summers, F. J., \& Davis, M. 1994, \apj, 422, 11

\item Gunn, J. E., \& Peterson, B. A. 1965, \apj, 142, 1633

\item Kang, H., Cen, R., Ostriker, J. P., \& Ryu, D. 1994, \apj, 428, 1

\item Katz, N., Hernquist, L., \& Weinberg, D. H. 1992, \apj, 399, L109

\item Little, B., Weinberg, D. H., \& Park, C. 1991, \mnras, 253, 295

\item Melott, A. L., \& Shandarin, S. F. 1990, \nat, 346, 633

\item Navarro, J. F., \& White, S. D. M. 1994, \mnras, 267, 401

\item Owen, J. M, Villumsen, J. V., Shapiro, P. R., \& Martel, H. 1996,
submitted \apjs\ December 1995

\item Shapiro, P. R., Giroux, M. L, \& Babul, A. 1994, \apj, 427, 25

\item Shapiro, P. R., \& Struck-Marcell, C. 1985, \apjs, 57, 205

\item Steinmetz, M., \& M\"{u}ller, E. 1994, \aap, 281, L97

\item Thoul, A. A., \& Weinberg, D. H. 1996, preprint

\item Weinberg, D. H., Hernquist, L., \& Katz, N. 1996, submitted to \apj,
astro-ph/9604175

\item White, S. D. M., \& Rees, M. J. 1978, \mnras, 183, 341
\end{description}

\clearpage
\figcaption{The ratio of the Jeans mass to the resolved mass ($M_J/M_R$) as
a function of expansion for each of the three resolutions used in this
paper ($N = 64^2, 128^2, 256^2$) for the $M_J > 0$ case.  The Jeans mass is
calculated using the average background density of the universe at each
expansion.
\label{MJ.fig}}

\figcaption{Dark matter overdensities
($\protect\Sub{\rho}{dm}/\bar{\rho}_{dm}$) for $M_J=0$ simulations.  Panels
arranged with increasing resolution along rows ($N = 64^2, 128^2, 256^2$),
and increasing cutoff in initial input perturbation spectrum down columns
($k_c = 32, 64, 128$).  Part a) shows results using the full resolution of
each simulation, while b) is calculated after resampling the simulations
down to equivalent $N=64^2$ resolution.  All simulations are shown at the
final time slice (expansion factor $a/a_i = 60$), with grey scale intensity
scaled logarithmically with dark matter density.
\label{DMRhoMaps.fig}}

\figcaption{Normalized dark matter overdensity distribution functions
$f(\protect\Sub{\rho}{dm}/\bar{\rho}_{dm})$ for $M_J=0$ (solid lines) and
$M_J>0$ (dotted lines) simulations.  Part a) shows results for full
simulations, while part b) shows all simulations degraded to equivalent
$N=64^2$ resolution.
\label{DMRhoDist.fig}}

\figcaption{Baryon overdensities
$\protect\Sub{\rho}{bary}/\protect\Sub{\bar{\rho}}{bary}$ for a) $M_J=0$ at
$a/a_i=30$, b) $M_J=0$ at $a/a_i=60$, c) $M_J>0$ at $a/a_i=30$, and d)
$M_J>0$ at $a/a_i=60$.  Panels arranged as in Figure
\protect\ref{DMRhoMaps.fig}.
\label{BaryRhoMaps.fig}}

\figcaption{Baryon overdensities for $k_c=32$ simulations, resampled to
$N=64^2$ resolution as in Figure \protect\ref{DMRhoMaps.fig}b.  Note these
panels represent the same simulations as the top rows of the previous
figures.  Shown are expansion factors a) $a/a_i=30$ and b) $a/a_i=60$.
Panels are arranged with increasing simulation resolution ($N = 64^2,
128^2, 256^2$) along rows, and increasing baryon Jeans mass ($M_J = 0$,
$M_J > 0$) down columns.
\label{BaryRhoMaps_R64.fig}}

\figcaption{Normalized baryon overdensity distribution functions 
$f(\protect\Sub{\rho}{bary}/\bar{\rho}_{bary})$ for $M_J=0$ (solid lines)
and $M_J>0$ (dotted lines).  We show the full resolution results for a)
$a/a_i=30$ and b) $a/a_i=60$, as well as results when each simulation is
degraded to $N=64^2$ resolution at c) $a/a_i=30$ and d) $a/a_i=60$.  Panels
arranged as in Figure \protect\ref{DMRhoDist.fig}.
\label{BaryRhoDist.fig}}

\figcaption{Kolmogorov-Smirnov statistic $D$ comparing
$f(\protect\Sub{\rho}{bary})$ between simulations.  Each line type
corresponds to one simulation which is compared to each simulation listed
on the ordinate axis, where the simulations are denoted as $N:k_c$.  Note
that the K-S statistic for comparing an individual simulation to itself is
formally $D=0$, but for the sake of clarity we have interpolated over these
points in this plot.  The panels are arranged with Jeans mass $M_J$
increasing along rows, and expansion $a/a_i$ increasing down columns.
\label{KS.fig}}

\figcaption{Baryon mass distribution for the $M_J=0$ simulations as a
function of overdensity and temperature
$f(\protect\Sub{\rho}{bary}/\protect\Sub{\bar{\rho}}{bary}, T)$ (upper row)
and $f(\protect\Sub{\rho}{bary}/\protect\Sub{\bar{\rho}}{bary}, T/T_{ad})$
(lower row).  $T_{ad} = T_0 (\rho/\rho_0)^{\gamma - 1}$ is defined as the
temperature the gas would have due solely to adiabatic processes.  Panels
arranged as in Figure \protect\ref{BaryRhoDist.fig}.
\label{RhoTDist.fig}}

\figcaption{Average baryon to dark matter mixture as a function of baryon
overdensity at $a/a_i=60$.  The baryon to dark matter mixture is defined as
$\protect\Sub{n}{bary}/\protect\Sub{n}{dm} = 
\protect\Sub{\Omega}{dm}\protect\Sub{\rho}{bary}/
\protect\Sub{\Omega}{bary}\protect\Sub{\rho}{dm}$, so that
$\protect\Sub{n}{bary}/\protect\Sub{n}{dm} > 1$ represents baryon enriched
material, while $\protect\Sub{n}{bary}/\protect\Sub{n}{dm} < 1$ is baryon
depleted.  In each panel the solid line shows the measured average baryon
to dark matter mixture, while the dashed lines represent the mixtures such
that 10\% of the mass at each overdensity is above and below the enclosed
region.  The dotted line shows the universal average
$\protect\Sub{n}{bary}/\protect\Sub{n}{dm} = 1$.  The top and bottom rows
represent the $M_J=0$ and $M_J>0$ simulations, respectively.
\label{B2D.fig}}

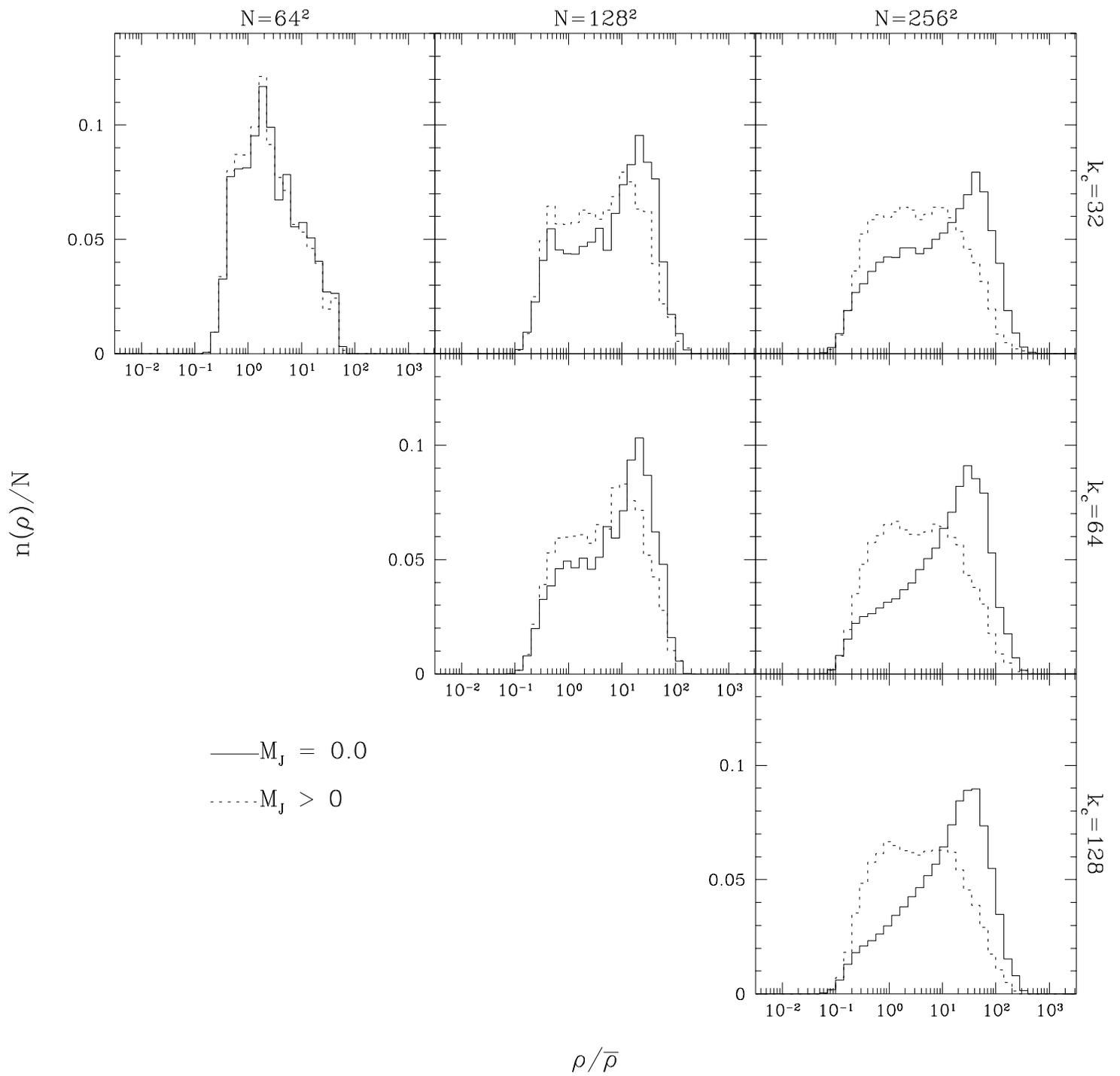
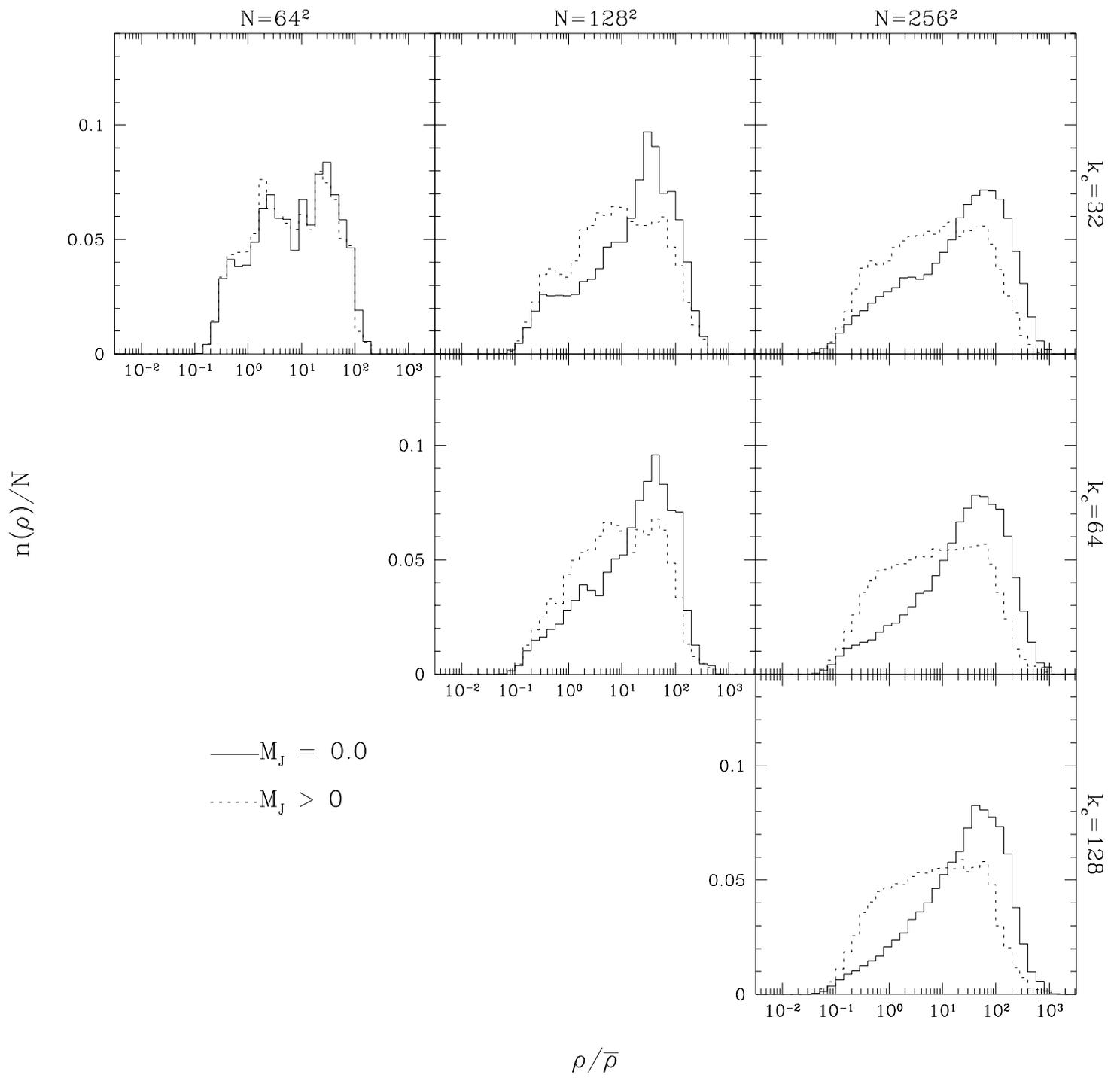
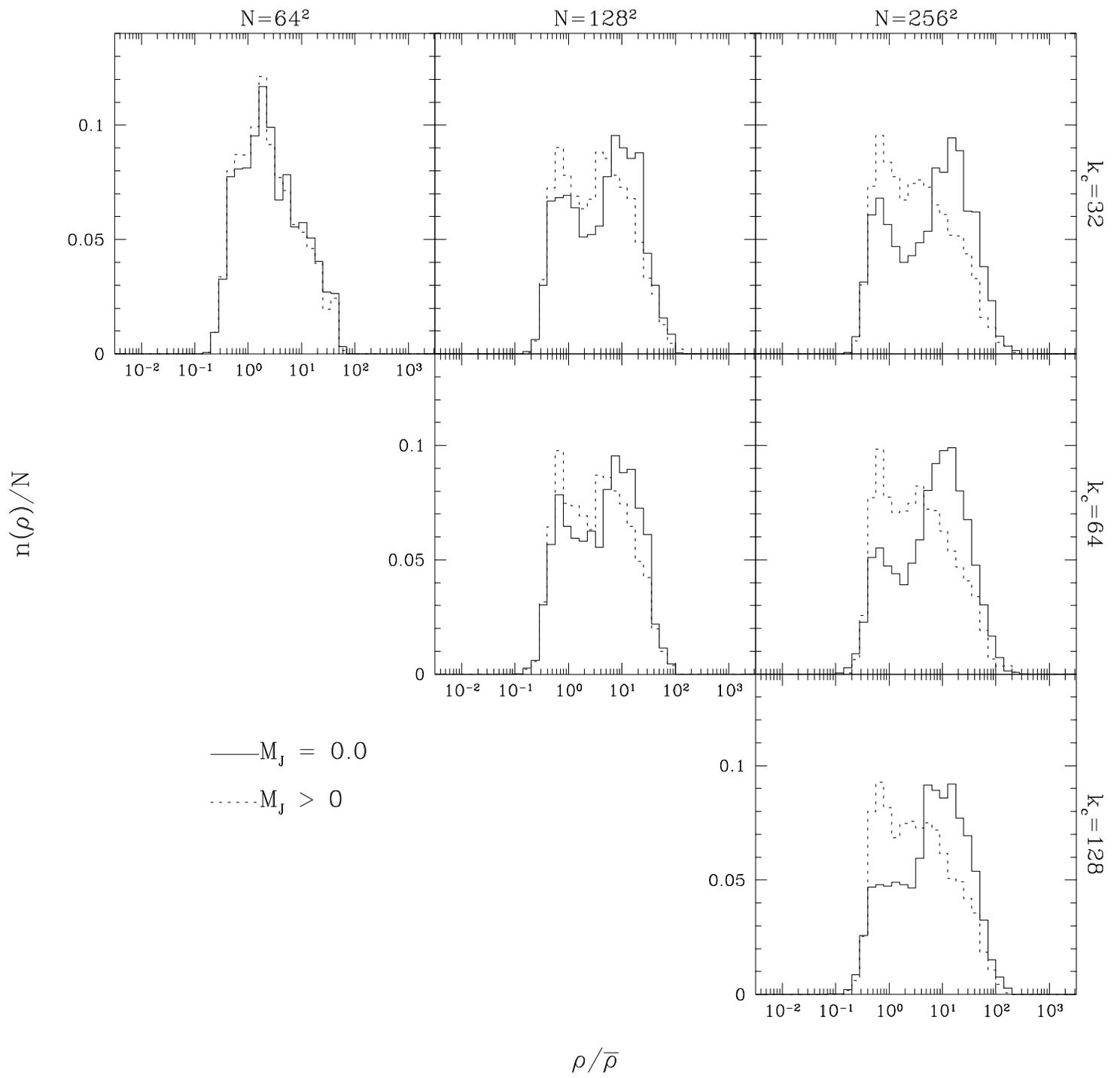
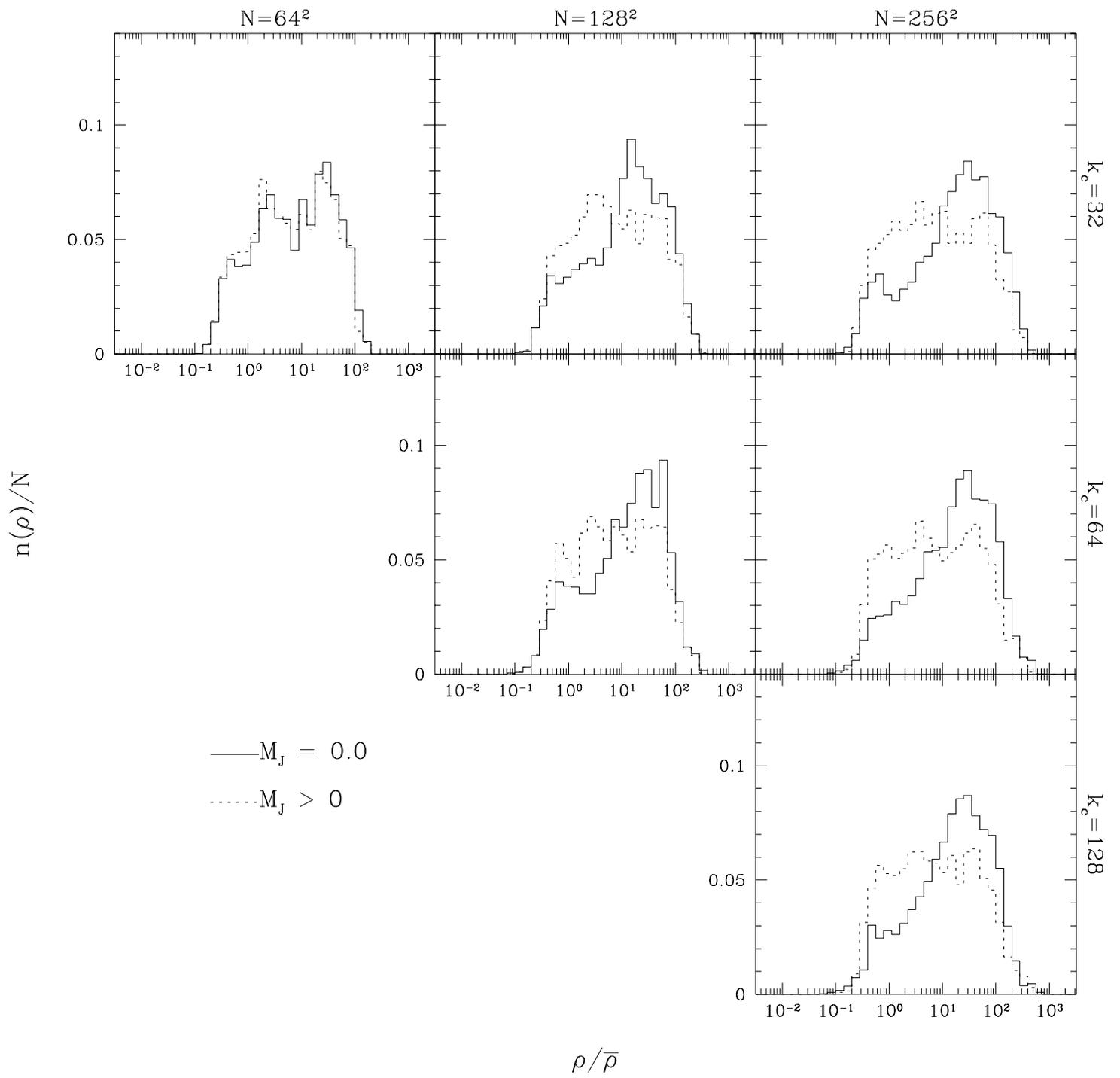
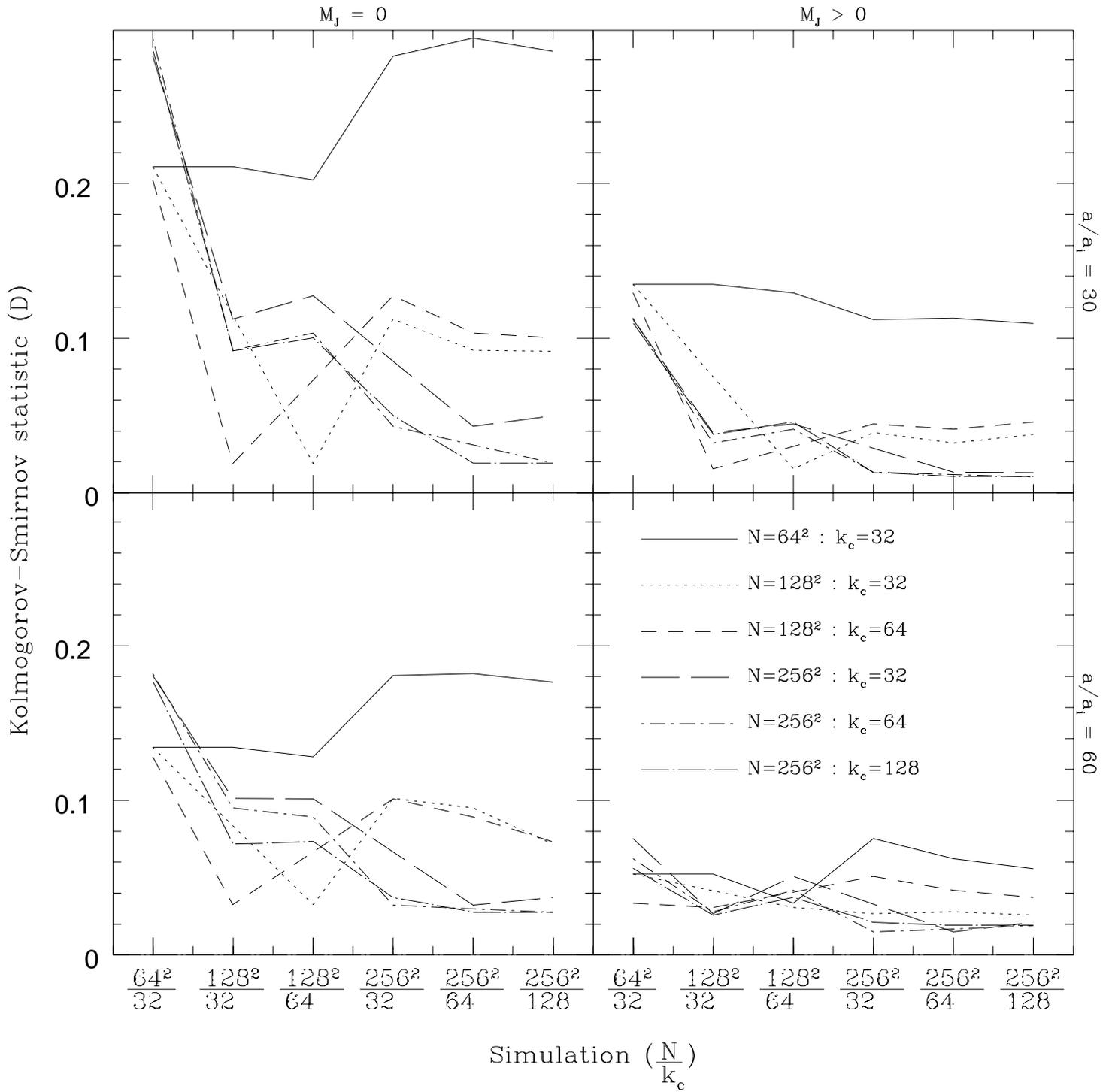
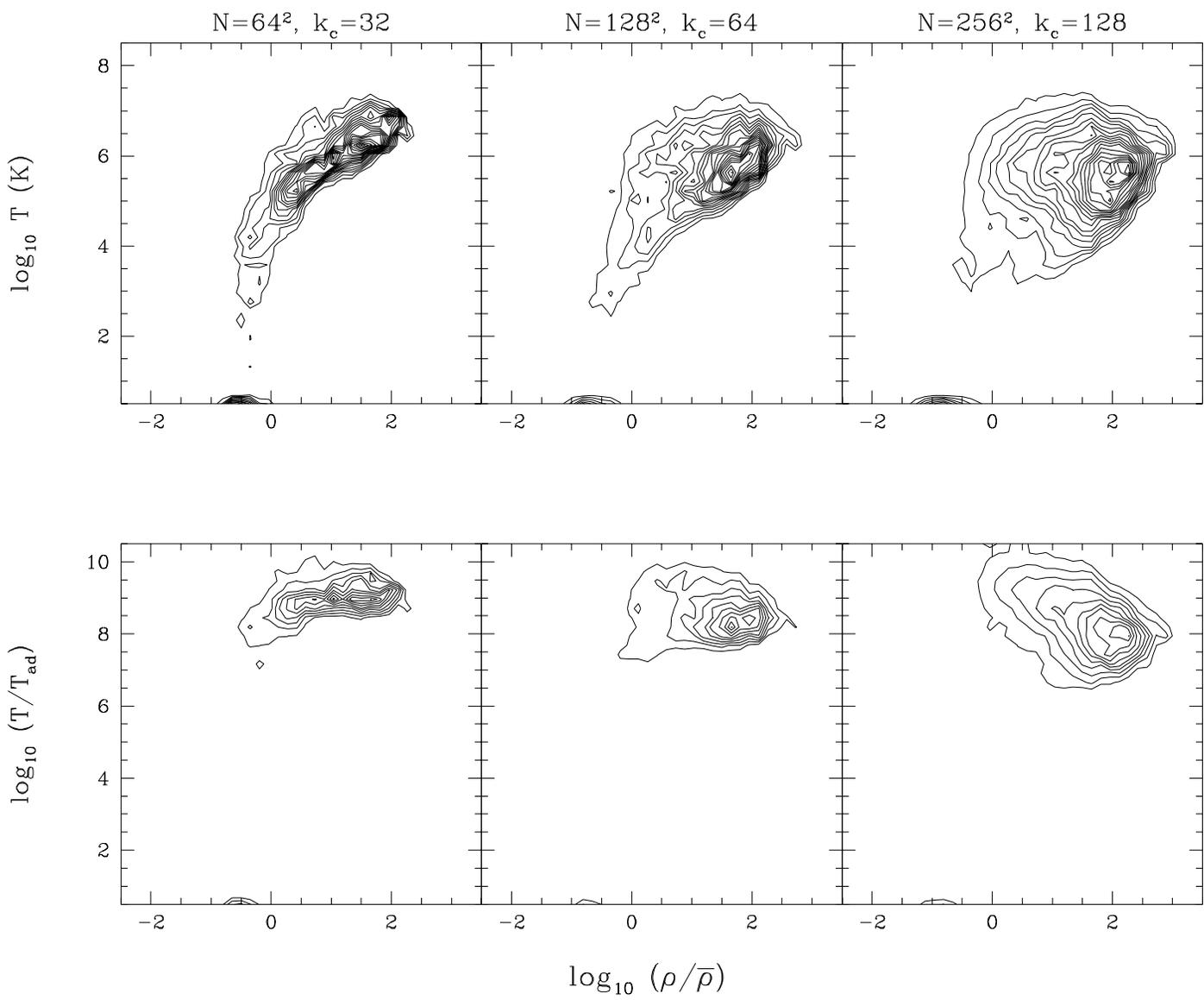
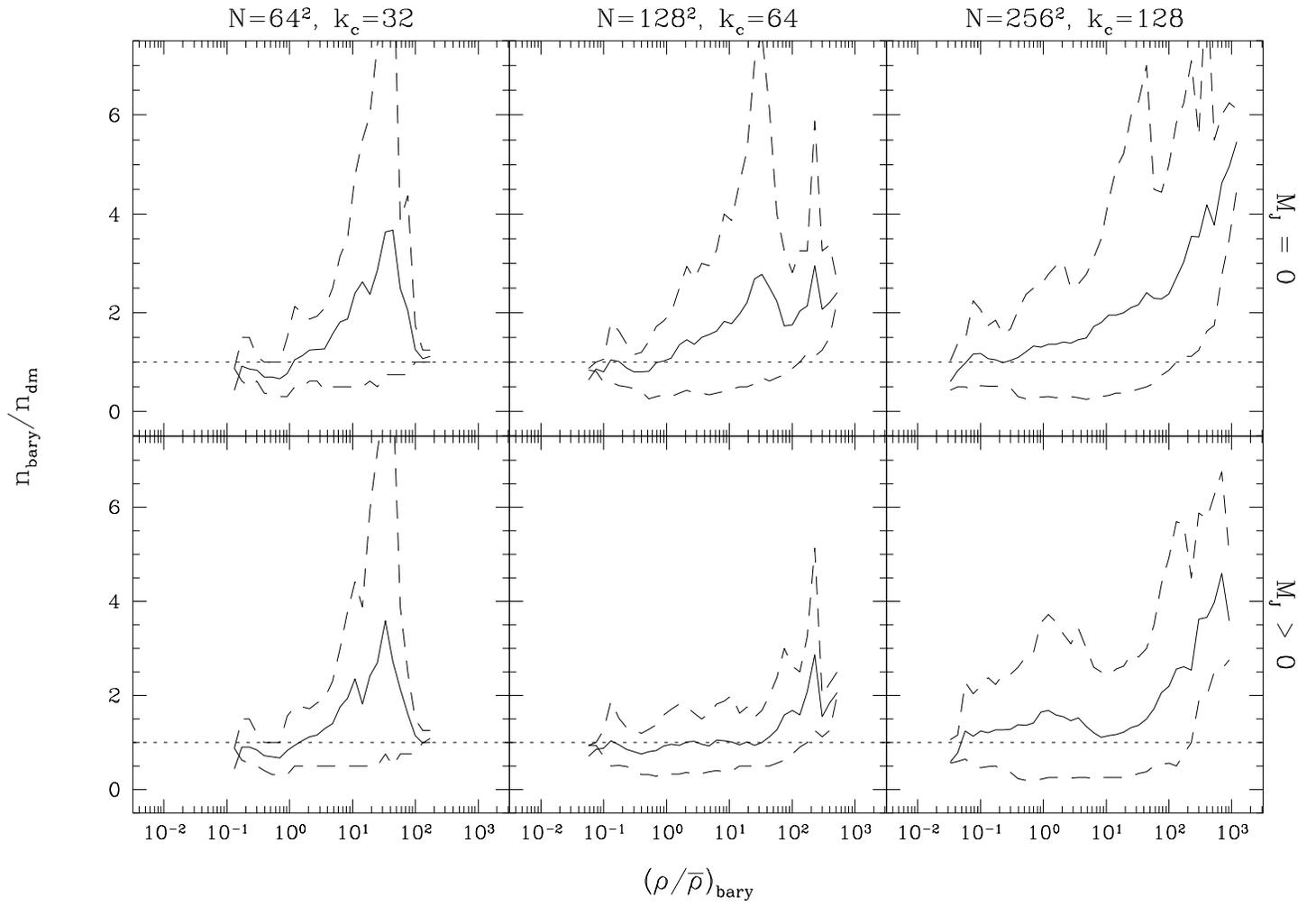
\end{document}